\documentclass[11pt,aps,nofootinbib,superscriptaddress,amssymb]{revtex4-1}

\usepackage{amsmath,setspace,subfigure,amsfonts,latexsym}
\usepackage{amssymb}
\usepackage{color}
\usepackage{epsfig}
\usepackage{graphicx}
\definecolor{White}{rgb}{1,1,1}
\definecolor{Red}{rgb}{1,0.1,0}
\definecolor{LightYellow}{rgb}{1,1,.875}
\definecolor{SteelBlue}{rgb}{.273,.508,.703}
\definecolor{navy}{rgb}{0,0,.5}
\definecolor{LightCyan}{rgb}{.875,1,1}
\definecolor{DarkRed}{rgb}{.543,0,0}
\definecolor{HotPink}{rgb}{1,.41,.70}
\definecolor{ForestGreen}{rgb}{.13,.54,.13}
\definecolor{OliveDrab}{rgb}{.42,.55,.14}
\definecolor{MediumBlue}{rgb}{0,0,.80}
\definecolor{RoyalBlue}{rgb}{.25,.41,.88}
\definecolor{DeepSkyBlue}{rgb}{0,.746,1}
\definecolor{Brown}{rgb}{0.545,0.271,0.074}

\def\bea{\begin{eqnarray}}
\def\eea{\end{eqnarray}}
\def\bec{\begin{center}}
\def\ec{\end{center}}

\def\beq{\begin{equation}}
\def\eeq{\end{equation}}

\newcommand\lsim{\mathrel{\rlap{\lower4pt\hbox{\hskip1pt$\sim$}}
    \raise1pt\hbox{$<$}}}
\newcommand\gsim{\mathrel{\rlap{\lower4pt\hbox{\hskip1pt$\sim$}}
    \raise1pt\hbox{$>$}}}
\def\bea{\begin{eqnarray}}
\def\eea{\end{eqnarray}}
\def\ba{\begin{array}}
\def\ea{\end{array}}
\def\bc{\begin{center}}
\def\ec{\end{center}}

\begin{document}

\title{\Large Light neutralino dark matter with a very light Higgs\\ 
for CoGeNT and DAMA/LIBRA data}

\author{Kyu Jung Bae\footnote{baekj81@phya.snu.ac.kr}, Hyung Do Kim\footnote{hdkim@phya.snu.ac.kr} and Seodong Shin\footnote{sshin@phya.snu.ac.kr}}

\affiliation{{\it FPRD and Department of Physics and Astronomy, Seoul National University, Seoul 151-747, Korea}}

\begin{abstract}
Recently, the CoGeNT collaboration reported the WIMP candidate signal events exceeding the known backgrounds where the light WIMP with large cross section is supported. Motivated by this issue, we analyze a light neutralino dark matter scenario with a very light CP-even Higgs mediation in the elastic scattering process, which provides the mass and direct detection cross section to explain the CoGeNT result. To be compatible with the result of LEP experiments, the light CP-even Higgs is favored to be in 9 to 10 GeV. Such a scenario can be realized in the ``Beyond the MSSM" context. The relic abundance consistent with the WMAP result can be obtained when twice of neutralino mass is close to the light Higgs mass via the resonance enhancement of the annihilation cross section. As a result, the neutralino mass is predicted to be at around 5 to 6 GeV.
\end{abstract}

\maketitle

\section{Introduction}
\label{sec:intro}

It was the measurement of velocity dispersions in the Coma cluster by Fritz Zwicky in 1933 \cite{Zwicky:1933gu} which led the requirement of nonluminous matter in our universe for the first time. Since then, a wide variety of evidence on nonbaryonic dark matter (DM) has been accumulated, such as the galactic rotation curve \cite{rotation} and the bullet cluster \cite{Bullet}. 

The current relic abundance of cold dark matter (CDM) in our universe has been observed by the WMAP for seven years \cite{WMAP} such that 
\begin{eqnarray}
0.1088 < \Omega_{\text{CDM}} h^2 < 0.1158,  \hspace{1cm} (1 \sigma \ \mbox{C.L.}) \label{eq:DMrelic}
\end{eqnarray}
where the scaled Hubble constant $h = 0.702^{+0.0013}_{-0.0014}$ in the units of $100$ km/sec/Mpc, combined with distance measurements from type Ia supernovae and baryon acoustic oscillations. This result indicates that about 23.3 \% of our universe, or 84.4 \% of matter is non-baryonic CDM, which motivates the theories beyond the Standard Model (SM). 

Various candidates of CDM explaining the observed relic abundance has been proposed. Among them, the most promising one is Weakly Interacting Massive Particle (WIMP) such as the lightest supersymmetric particle (LSP) in the supersymmetric models with R-parity \cite{LSP1, LSP2}, the lightest Kaluza-Klein (KK) particle in the extra dimensional models with KK parity \cite{LKKP}, the lightest T-odd particle in the T-parity conserved little Higgs model \cite{Todd}, and the SM gauge singlet particles in Higgs portal models \cite{SDM}. WIMPs are produced at the early stage of our universe and their current relic abundance is naturally determined when their interactions to the SM particles freeze out \cite{Lee:1977ua}.

In the meantime, direct detection experiments to detect WIMP scattering off target nuclei have been constructed to figure out the physical properties of them. Recently, the Coherent Germanium Neutrino Technology (CoGeNT) experiment reported that about a hundred events \cite{cogent} exceeding the expected background are observed after their eight week operation, which possibly originated from the nuclear recoil by DM scattering. Due to its enhanced sensitivity to low energy events, the ionization signal which is observed in the CoGeNT detector is as low as $0.4 - 3.2$ keVee\footnote{The ``ee" denotes ``electron equivalent", which indicates the kinetic energy of the electron inducing the ionization.}. Comparing with the energy threshold of DAMA/LIBRA \cite{dama} which is $2$ keVee combined with the channeling effect, this energy threshold is tremendous. The discovery region, hence, supports the existence of light DM whose mass is $m_{\chi} \lsim 10$ GeV and the spin independent WIMP-nucleon elastic scattering cross section $\sigma_{\text{SI}}$ is as high as $\sim 10^{-40}$ $\text{cm}^2$.  

The light WIMP with large cross section is also favored in the recent DAMA/LIBRA annual modulation signal. Considering the channeling effect \cite{channeling}, the result allows sizable parameter space compatible with other null experiments. There have been several proposals to explain this, such as light SM gauge singlet fermion \cite{SFDMdama} or scalar DM \cite{andreas}, the WIMPless model \cite{wimpless}, light neutralinos in the minimal supersymmetric standard model (MSSM) without the assumption of gaugino mass unification at the grand unified theory (GUT) scale \cite{bottino}, right-handed sneutrino DM in an extended model of MSSM \cite{cerdeno}, and the mirror DM model \cite{foot}, etc. Therefore, it is very interesting to propose a plausible light WIMP scenario\footnote{It must be noted that there exists some negative research positing that the channeling fraction of recoiling lattice nuclei in NaI is quite suppressed to provide its meaningful effect \cite{Bozorgnia:2010xy}.}.
 
In this paper, we analyze a light neutralino DM scenario since  supersymmetry (SUSY) is the most promising candidate among the theories beyond the SM. In order to have such a large scattering cross section in a supersymmetric SM, either large value of $\tan\beta$ or a very light CP-even Higgs boson mediator is needed in the process of elastic scattering of the neutralino off the target nuclei. Such scenarios cannot be easily realized in the context of the MSSM since they are highly constrained by other experiments such as the LEP, Tevatron, and rare decays. Especially, it is not easy to obtain $\sigma_{\text{SI}}$ as high as $\sim 10^{-40}\text{ cm}^2$ for the explanation of the CoGeNT result with conventional halo parameters, even though we consider the previous approach to explain the DAMA/LIBRA result such as the reference \cite{bottino}\footnote{The experimental bounds from the neutral Higgs bosons $\to \tau^+ \tau^-$ and the rare decay $B \to \tau \nu$ are needed to be considered in \cite{bottino}.}. Therefore, the ``Beyond the MSSM" (BMSSM) considerations are required in this paper. Without considering additional light degrees of freedom, the light Higgs scenario is natural to be studied first, which we focus on here. (Scenarios with light degrees of freedom are explained in \cite{lightnmssm} with the NMSSM context and in \cite{SFDMdama} with the SM gauge singlet Dirac fermion. The latter can be also obtained by slightly changing the NMSSM potential with singlet quadratic terms.) We start from analyzing the parameter space to obtain the large detection cross section $\sigma_{\text{SI}} \sim 10^{-40}\text{ cm}^2$ for the light WIMP of $4$ GeV $\lsim m_{\chi} \lsim$ $7$ GeV in order to avoid the bounds from the null experiments. Due to the limits by the LEP experiments and $\Upsilon$ decay, the parameter space of light Higgs is constrained to be in the window of 9 to 10 GeV. It is, then, interesting that the resonance channel in the WIMP annihilation process is naturally considered on the viable parameter domain, which provides the right relic abundance of the light neutralino. Therefore, the light neutralino with mass of 5 to 6 GeV can explain the correct relic abundance. It must be noticed that this research does not concern the details of the theoretical construction but will focus on the compatible parameter space in the supersymmetric model to explain the viable light WIMP consistent with the various experiments.  

\begin{figure}
\begin{center}
\includegraphics[width=14cm]{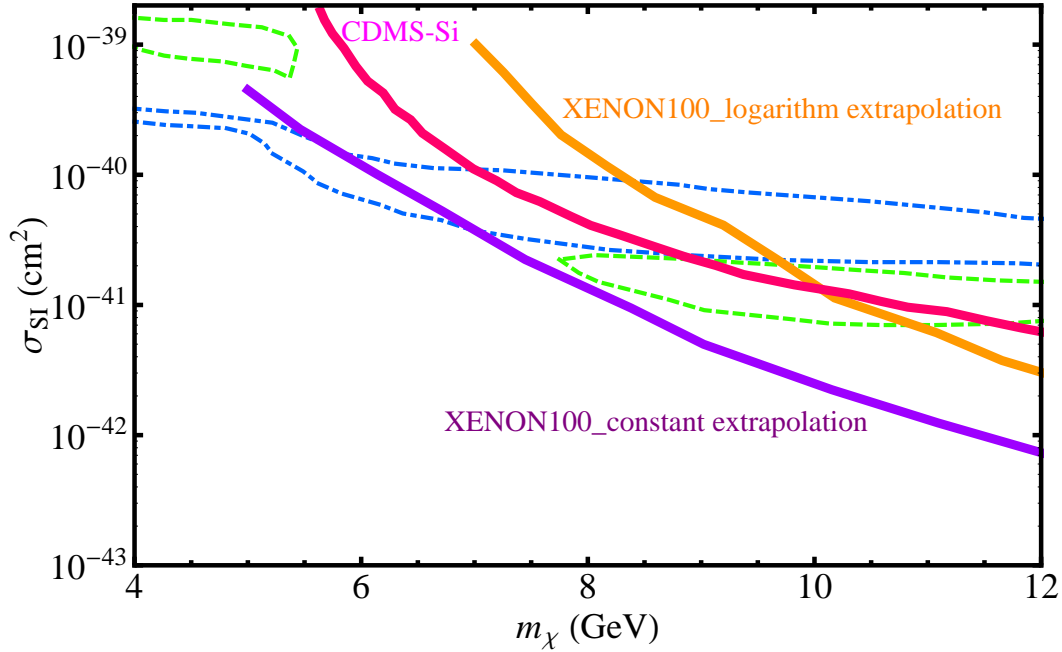}
\end{center}
\caption{Spin independent elastic scattering $\sigma_{\text{SI}}$ on the WIMP mass $m_{\chi}$ is plotted. The halo profile is given as $v_0 \approx 270$ km/s and $v_{\text{esc}} \approx 490$ km/s. This figure contains the exclusion limit by the recent XENON100 experiment. $\mathcal{L}_{\text{eff}}$ at low energy is assumed by the constant extrapolation (violet line) or logarithm extrapolation (orange line) from the global fit and lower contour of $\mathcal{L}_{\text{eff}}$. (See Fig. 3 of \cite{xenonreply}.) The limit by the CDMS-Si (magenta line) was the strongest one for light WIMPs before the the XENON100 eleven day result (with constant extrapolation of $\mathcal{L}_{\text{eff}}$). The constant extrapolation of $\mathcal{L}_{\text{eff}}$, however, seems to be a too naive assumption. The papers \cite{collar:2010} introduce another measurement of $\mathcal{L}_{\text{eff}}$ which shows a drastic decrease below $10$ keV nuclear recoil energy \cite{Lebedenko:2008gb}. The annual modulation signals observed in the DAMA/LIBRA experiment are inside the green dashed line, where the channeling effect is considered and the lowest energy bin is neglected. The recent CoGeNT result is inside the blue dot dashed line considering the 50 \% exponential background contribution.}
\label{fig:si}
\end{figure}

There have been researches to reconcile the CoGeNT report on WIMP signals with other null experiments constraining the detection bound and the previous DAMA result \cite{cogentpapers1,cogentpapers2}. At first sight, it seems that the CoGeNT signal events are excluded by other experiments mainly due to the CDMS-Si \cite{cdmssi} and the recent XENON100 \cite{xenon100}. It is important, however, that the CoGeNT collaboration has not analyzed all the background events such as surface events through the intra contact surface and those by electro-formed cryostat components, etc. It is, hence, possible for the sizable CoGeNT signals to avoid the exclusion limits by considering proper background models, combined with changing the WIMP distribution or detector parameters such as halo profile and scintillation efficiency\footnote{The fraction of the nuclear recoil energy to scintillation energy. Recent result is in Fig.1 of \cite{xenonreply}.} $\mathcal{L}_{\text{eff}}$ used in those experiments. After the recent XENON100 collaboration claimed to exclude the CoGeNT and DAMA signal candidate region, a lot of hot discussions have been made \cite{xenonreply,collar:2010}. They, however, seem to agree that the $\mathcal{L}_{\text{eff}}$ below 10 keV nuclear recoil energy has ambiguities so that some CoGeNT region can survive even without consideration of proper background. Considering those arguments, we show the allowed signal region of the CoGeNT and DAMA with proper background models in Fig. \ref{fig:si}. $\mathcal{L}_{\text{eff}}$ at low energy is assumed to be determined by the constant extrapolation (violet line) from the global fit or logarithm extrapolation (orange line) from the lower contour of $\mathcal{L}_{\text{eff}}$ in the XENON100 experiment. The constant extrapolation of $\mathcal{L}_{\text{eff}}$ seems to be a too naive assumption. The papers \cite{collar:2010} introduce another measurement of $\mathcal{L}_{\text{eff}}$ which shows a drastic decrease below $10$ keV nuclear recoil energy \cite{Lebedenko:2008gb}. The recent CoGeNT result is inside the blue dot dashed line considering the 50 \% exponential background contribution, which covers lower mass region $4$ GeV $\lsim m_{\chi} \lsim 7$ GeV \cite{cogentpapers2}. Meanwhile, the annual modulation signals observed in the DAMA/LIBRA experiment considering the channeling effect are shown inside the green dashed line. In this analysis, the lowest energy bin is neglected so that the signal candidate region is broader than those in \cite{xenon100,xenonreply}. Reconciling the CoGeNT result with the DAMA/LIBRA result is also possible by changing the fraction of channeling effect in DAMA/LIBRA to lower values than expected. 
Consequently, the light WIMP of mass $4 - 7$ GeV with the spin independent elastic cross section $\sigma_{\text{SI}} \sim 10^{-40}\mbox{ cm}^2$ is the plausible candidate to explain the CoGeNT and DAMA/LIBRA result\footnote{Additional situation after the submission of the first version of this paper will be explained in our Sec. \ref{sec:relic}.}. Therefore, we focus on the reasonable parameter space in the BMSSM. 

The paper is organized as follows. We discuss the search for a light neutralino in the context of ordinary MSSM in Sec. \ref{sec:cogentandmssm}. Experimental constraints on the mass of a light neutralino are covered in Sec. \ref{sec:lightneutralino}. A very light Higgs to explain the direct detection, simultaneously considering the LEP bounds is discussed in Sec. \ref{sec:lightHiggs}. The light Higgs scenario in the BMSSM will be discussed with numerical results in Sec. \ref{sec:lightBMSSM}. We also discuss the stability of the BMSSM scalar potential in Sec. \ref{sec:stability}. The relic abundance of the neutralino is computed in Sec. \ref{sec:relic}. Finally, we conclude in Sec. \ref{sec:conclusions}.

\section{WIMP-nucleon scattering in the MSSM}
\label{sec:cogentandmssm}

The spin independent elastic scattering cross section of the neutralino-nucleon system is given by \cite{LSP1,LSP2}
\begin{equation}
\sigma = \frac{4m_r^2}{\pi}f_{p(n)}^2~,
\end{equation}
where $m_r$ is the reduced mass of the neutralino-nucleon system and  
$f_{p(n)}$ is the neutralino coupling to a proton (neutron) which is composed of i) neutralino-(light) quark current and ii) neutralino-gluon couplings such that, in large squark mass limit \cite{LSP1}, 
\begin{equation}
f_{p,n}=\sum_{q=u,d,s}f_{T_q}^{(p,n)}a_q\frac{m_{p,n}}{m_q}
+\frac{2}{27}f_{TG}^{(p,n)}\sum_{q=c,b,t}a_q\frac{m_{p,n}}{m_q} ,
\label{eq:fp}
\end{equation}
where   
$f_{T_u}^{(p)}=0.020\pm0.004$, $f_{T_d}^{(p)}=0.026\pm0.005$, $f_{T_s}^{(p)}=0.118\pm0.062$,
$f_{T_u}^{(n)}=0.014\pm0.003$, $f_{T_d}^{(n)}=0.036\pm0.008$ and $f_{T_s}^{(n)}=0.118\pm0.062$ \cite{Ellis}, and 
\begin{equation}
f_{TG}^{(p,n)} = 1-\sum_{q=u,d,s}f_{T_q}^{(p,n)} .
\end{equation}

Meanwhile, the neutralino-quark coupling $a_q$ which is defined in the effective Lagrangian as $a_q \bar{\chi} \chi \bar{q}q$, is obtained by
\begin{equation}
a_q = \sum_{\phi=h,H}\frac{g_{\phi\chi\chi}g_{\phi qq}}{2m_{\phi}^2} ,
\end{equation}
where the couplings $g_{\phi\chi\chi}$ and $g_{\phi qq}$ are given in the following Feynman rules
(\ref{feyn:hNN}) and (\ref{feyn:HNN})\footnote{The couplings are defined in the Lagrangian such that $\frac12 g_{\phi \chi \chi} \bar{\chi} \chi \phi$ and $g_{\phi q q} \bar{q} \phi q$.}.
Here we neglect squark-mediated contribution since squarks are usually much heavier than Higgs bosons to satisfy collider search bounds.
Feynman rules for Higgs couplings to neutralinos are adopted from \cite{LSP1,Drees:2004jm} such that
\begin{eqnarray}
& & g_{h\chi\chi}=-g_2(Q_{11}^{''}\sin\alpha+S_{11}^{''}\cos\alpha),\label{feyn:hNN}\\
& & g_{H\chi\chi}=-g_2(-Q_{11}^{''}\cos\alpha+S_{11}^{''}\sin\alpha),\label{feyn:HNN}\\
& & g_{A\chi\chi}=-ig_2(Q_{11}^{''}\sin\beta-S_{11}^{''}\cos\beta),\label{feyn:ANN}\\
& & g_{huu} = \frac{m_u}{\sqrt{2}v}\frac{\cos\alpha}{\sin\beta},\quad
g_{Huu} = \frac{m_u}{\sqrt{2}v}\frac{\sin\alpha}{\sin\beta},\quad
g_{Auu} = i\frac{m_u}{\sqrt{2}v}\cot\beta \gamma_5,\label{feyn:huu}\\
& & g_{hdd} = -\frac{m_d}{\sqrt{2}v}\frac{\sin\alpha}{\cos\beta},\quad
g_{Hdd} = \frac{m_d}{\sqrt{2}v}\frac{\cos\alpha}{\cos\beta},\quad
g_{Add} = i\frac{m_d}{\sqrt{2}v}\tan\beta\gamma_5,\label{feyn:hdd}
\end{eqnarray}
and 
\begin{eqnarray}
Q_{11}^{''}&=&N_{13}(N_{12}-\tan\theta_WN_{11}),\\
S_{11}^{''}&=&N_{14}(N_{12}-\tan\theta_WN_{11}),
\end{eqnarray}
where $N_{ij}$ is a neutralino mixing matrix in the basis of $(\tilde{B}, \tilde{W}^0, \tilde{H_d}^0, \tilde{H_u}^0)$, $v=174$ GeV is the Higgs vacuum expectation value,
$\tan\beta$ is the ratio of vacuum expectation values of two Higgses,
$\alpha$ is a CP-even Higgs mixing angle,
$\theta_W$ is a weak mixing angle, $u$($d$) stands for up(down)-type quarks, and $g_2$ is the gauge coupling of $SU(2)_L$ in the SM.
Keeping only the dominant contributions, we obtain the elastic WIMP - nucleon (N) scattering such that 
\begin{eqnarray}
\sigma &=& \frac{4m_r^2}{\pi} \left\{ f_{T_s}\left(\frac{m_N}{m_s}\right)
\left(\frac{g_{h\chi\chi} g_{hss}}{2m_h^2}
+\frac{g_{H\chi\chi}g_{Hss}}{2m_H^2}\right) \right. \nonumber \\
& & \hspace{1cm} \left. +~ \frac{2}{27} f_{TG} \left(\frac{m_N}{m_{q=c,b,t}}\right)
\left(\frac{g_{h\chi\chi} g_{hqq}}{2m_h^2}
+\frac{g_{H\chi\chi}g_{Hqq}}{2m_H^2}\right) \right\}^2~. \label{cross}
\end{eqnarray}
In the MSSM decoupling limit, $\beta-\alpha\simeq\pi/2$ and light Higgs $h$ is mostly up-type and SM-like
so that $g_{hqq}$ becomes SM one, {\it i.e.}, $g_{hqq}=m_q/(\sqrt{2}v)$,
but heavy Higgs $H$ becomes mostly down-type so that
$g_{Huu}\simeq0$, $g_{Hdd}\simeq m_d\tan\beta/(\sqrt{2}v)$.
Therefore, the scattering cross section for large $\tan\beta$ becomes
\begin{equation}
\sigma\simeq\frac{m_r^2 m_N^2}{\pi} \frac{1}{2v^2} \left(f_{T_s} + \frac{2}{27}f_{TG}\right)^2\biggl(
\frac{g_1N_{14}N_{11}}{m_h^2}+\frac{g_1N_{13}N_{11}}{m_H^2}\tan\beta\biggr)^2~,
\end{equation}
where $g_1=g_2\tan\theta_W$ is $U(1)_Y$ gauge coupling and we omit the $N_{12}$ contribution because it is negligible in typical MSSM parameter space \cite{Bae:2007pa}.
In the regime of very large $\tan\beta$ with $m_H$ not much larger than $m_h$, the neutralino-nucleon scattering process is dominated by the heavy CP-even Higgs mediated contribution so that the elastic scattering cross section becomes approximately
\begin{equation}
\sigma \simeq 0.23 \times 10^{-40}\text{ cm}^2\times\biggl(\frac{N_{13}}{0.4}\biggr)^2
\biggl(\frac{\tan\beta}{50}\biggr)^2\biggl(\frac{100\text{ GeV}}{m_H}\biggr)^4~, \label{eq:detect}
\end{equation}
for $m_{\chi} \sim 7$ GeV where the subdominant down quark and one-loop induced bottom quark contributions are also considered. Therefore, we need very large $\tan\beta > 100$ for $m_H=100$ GeV to explain the CoGeNT result. In the regime of such large $\tan\beta$, however, the branching ratio of $B_s \to \mu^{+}\mu^{-}$ severely constrains the parameter set. In addition, combining the upper limit on the neutral Higgs bosons $\to \tau^{+} \tau^{-}$ in the Tevatron and the observations of the rare decays $B \to \tau \nu$ with the ratio of $B \to D \tau \nu / B \to D l \nu$ in the B factories, we obtain the constraints on the elastic scattering $\sigma_{\text{SI}} \lesssim 5 \times 10^{-42} \text{cm}^2$, which is much lower than $10^{-40} \text{cm}^2$ to explain the CoGeNT result \cite{Feldman:2010ke,Kuflik:2010ah}. In the heavy $H$ scenario, it is hence impossible to construct viable models which support the light WIMP of mass $4 - 7$ GeV with the spin independent elastic cross section $\sigma_{\text{SI}} \sim 10^{-40}\mbox{ cm}^2$. In order to reduce low energy constraints, we need to invoke ``wrong-Higgs" interactions \cite{Bae:2010ai}.

\section{Experimental constraints on the mass of light neutralino}
\label{sec:lightneutralino}

The experimental constraints on the mass of a light neutralino are analyzed in \cite{Dreiner:2009ic} and the references therein.
In this section, we briefly review the experimental constraints which are relevant to our case. 
The most stringent constraints are produced by the LEP experiments.
The assumption of the gaugino mass unification at the GUT scale induces the relation of the $U(1)_Y$ gaugino mass parameter $M_1$ and the $SU(2)$ gaugino mass parameter $M_2$ at the electroweak scale such that $M_1 \approx \frac12 M_2$. Therefore, the bound of lightest neutralino mass can be obtained from the direct search on the chargino mass bound, $m_{\tilde{\chi}^{\pm}_1}>94$ GeV, which directly constrains $M_2$ and the Higgsino mass parameter $\mu$ and gaugino mass unification condition relates this bound to the light neutralino,  
\begin{equation}
m_{\tilde{\chi}_1^0}>46 \text{ GeV} .
\end{equation}
However, the above neutralino mass bound can be evaded if we are free from the assumption of gaugino mass unification at the GUT scale. Thus other experimental constraints must be considered to obtain the direct mass limit of light neutralino in the MSSM. The two kinds of constraints are very important for such discussion. One is from the invisible $Z$ boson decay width and the other is from the lightest - second lightest neutralino pair production,
$e^+e^-\to\tilde{\chi}_1^0\tilde{\chi}_2^0$ at the LEP2 experiment for the case $m_{\chi_1^0}+m_{\chi_2^0}<208$ GeV.

\begin{figure}
\begin{center}
\includegraphics[width=4cm]{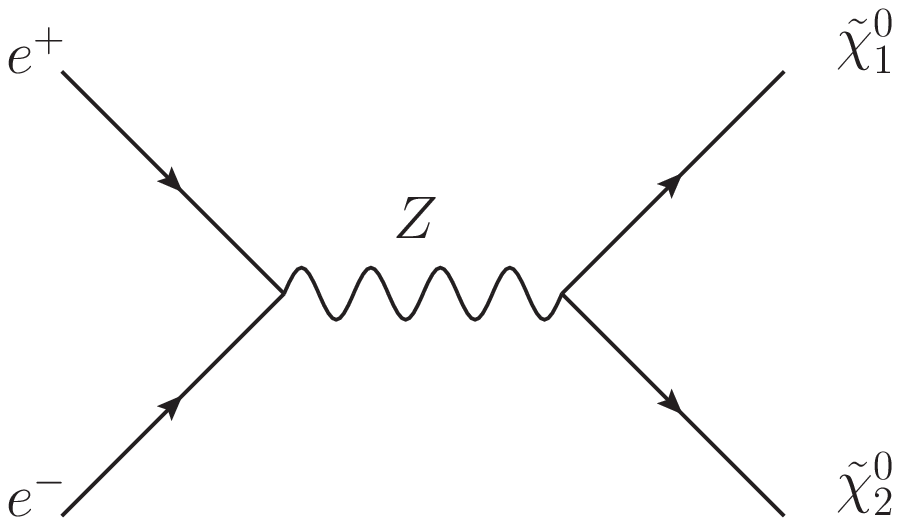}\qquad
\includegraphics[width=3.4cm]{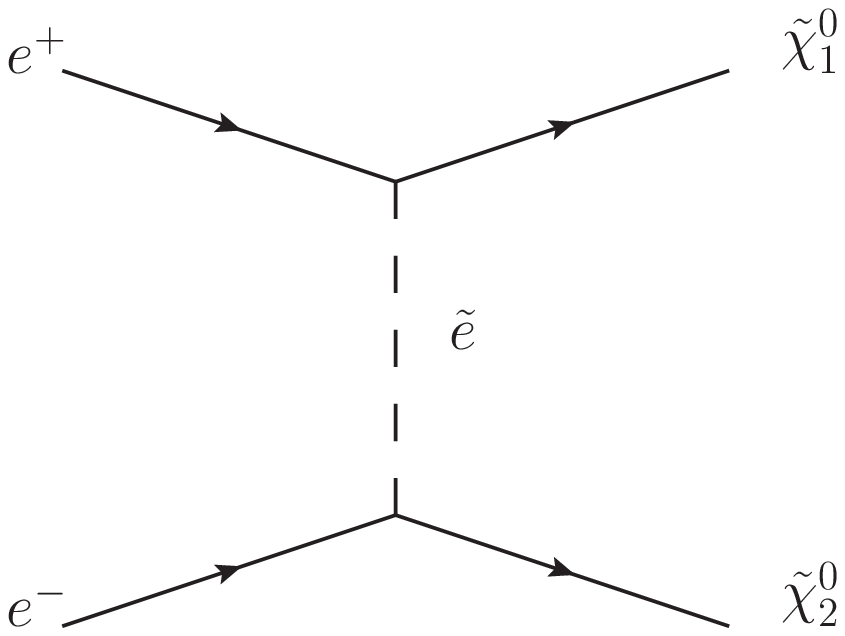}\qquad
\includegraphics[width=3.4cm]{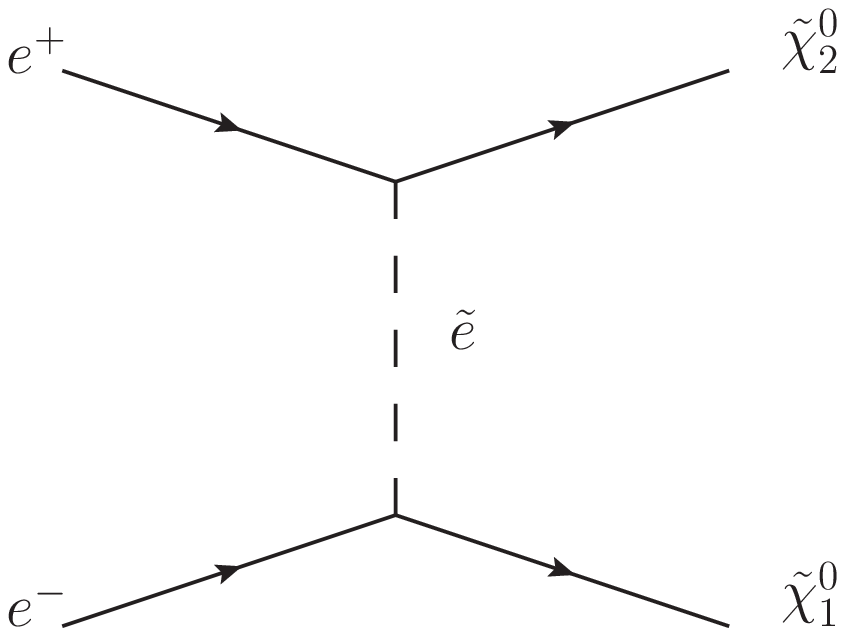}\qquad
\end{center}
\caption{Feynman diagrams for $e^+e^-\to\tilde{\chi}^0_1\tilde{\chi}^0_2$.}\label{neut_assoc}
\end{figure}

The invisible $Z$ decay width is described by \cite{Kuflik:2010ah},
\begin{equation}
\Gamma(Z\to\tilde{\chi}_1^0\tilde{\chi}_1^0)=\frac{g_2^2}{4\pi}\frac{(N_{13}^2-N_{14}^2)^2}{24\cos^2 \theta_W}
M_Z\biggl[1-\biggl(\frac{2m_{\tilde{\chi}_1^0}}{M_Z}\biggr)^2 \biggr]^{3/2},
\end{equation}
where $M_Z$ is the $Z$ boson mass and $\theta_W$ is the weak mixing angle.
The 2$\sigma$ experimental constraint, $\Gamma_{\text{inv}}<3$ MeV implies that $|N_{13}^2-N_{14}^2|\lsim0.13$.
As will be seen in the following sections, the parameter region of our light neutralino is obtained for $N_{13}\sim0.3$ and $N_{14}\sim0$, with 
$\mu=200$ GeV and $M_2=400$ GeV so that the mass of our light neutralino has no constraint from the $Z$ boson invisible decay width. As the other important consideration, the conservative bound for the lightest - second lightest neutralino pair production at the LEP2 is given by \cite{Dreiner:2009ic}
\begin{equation}
\sigma(e^+e^-\to\tilde{\chi}_1^0\tilde{\chi}_2^0)<70\text{ fb} ,
\end{equation}
where Br$(\tilde{\chi}_2^0 \to \tilde{\chi}_1^0 Z)=1$ is conservatively assumed.
This bound can be also evaded if the mass of scalar electron is larger than 500 GeV  for small $\mu$ \cite{Dreiner:2009ic} 
since scalar electron mediated neutralino production processes are suppressed by heavy scalar electron mass.
Feynman diagrams for this process are shown in Fig. \ref{neut_assoc}.
For large $\mu \gtrsim 200$ GeV, {\it i.e.} $m_{\chi_1}+m_{\chi_2}>208$ GeV, 
this process is kinematically disallowed so that the mass of scalar electron for neutralino production is not bounded at all. 
In addition to these constraints, the rare meson decays can also constrain the very light neutralino, however, these are again irrelevant since the branching ratios are much lower than the experimental bounds \cite{Dreiner:2009ic}. 
Consequently, the mass of our light neutralino to explain the CoGeNT and DAMA/LIBRA results is not constrained by the collider bounds, precision observables, and rare meson decays.

\section{Light Higgs Scenario and Experimental Constraints}
\label{sec:lightHiggs}

Instead of the heavy CP-even Higgs mediation scenario for very large $\tan\beta$, we consider here the case that the light CP-even Higgs mediated contribution dominates the elastic scattering with moderate $\tan\beta\simeq3$. Observing the eq. (\ref{cross}), light Higgs mediation can be
more important than that of heavy Higgs if $m_h \ll m_H$. In such a case, light Higgs mass $m_h$ is very small to have large cross section for neutralino-nucleon elastic scattering 
so that constraints from LEP experiments and rare decays of mesons must be considered. 

If $10$ GeV$\lsim m_h\lsim20$ GeV, two kinds of Higgs search at the LEP experiments must be under consideration. One is the Higgsstrahlung process, $e^+e^-\to Z^*\to hZ$,
and the other is the associative production, $e^+e^-\to Z^* \to hA$.
For convenience, we define the following quantities
\begin{eqnarray}
R_{hZ}&\equiv&
\frac{\sigma(e^+e^-\to Z^*\to Zh)_\text{MSSM}}{\sigma(e^+e^-\to Z^*\to Zh)_{\text{SM}}} 
=\sin^2(\alpha-\beta) , \\
& & \nonumber \\
R_{hA}&\equiv&\frac{\sigma(e^+e^-\to Z^*\to hA)_\text{MSSM}{\cal B}(h\to \bar{b}b){\cal B}(A\to \bar{b}b)}
{\sigma(e^+e^-\to Z^*\to hA)_\text{ref}}\\
&=&\cos^2(\alpha-\beta){\cal B}(h\to \bar{b}b){\cal B}(A\to \bar{b}b) ,\nonumber
\end{eqnarray}
where $\sigma(e^+e^-\to Z^*\to hA)_\text{ref}$ is a reference value assuming that 
$Z-h-A$ coupling constant is equal to that of the SM $Z-Z-h$ coupling, {\it i.e.}, $g_{ZhA}=g_{ZZh}^{\text{SM}}$.
In order to satisfy the negative results of scalar searches at the LEP experiments, $R_{hZ}\lsim0.01$ is required for the Higgsstrahlung process and $R_{hA}\lsim0.2$ for the associative production when $m_h\sim20$ GeV, $m_A\sim 90$ GeV \cite{LHWG-Note:2005-01}.

In the case that $\sin^2(\alpha-\beta)<0.01$, we can evade the constraint from the Higgsstrahlung process.
However, avoiding the associative production constraint is not trivial since $\cos(\alpha-\beta)\simeq1$.
The light neutralino with $m_{\chi}\lsim m_h /2$ can be a solution in this case. Since the light Higgs can decay to the neutralinos, the branching ratio of Higgs decay to neutralinos is comparable to or larger than that of Higgs decay to $b$-quark pair for low $\tan\beta\lesssim3$ \cite{Yaguna:2007vm}. Consequently, we have a reliable parameter space by constraining $m_{h} \gtrsim 2 m_{\chi}$ when $10$ GeV $\lsim m_h \lsim$ $20$ GeV. It is, however, not the end of the story and this region will be discussed again after our eq. (\ref{eq:pmatrix}).

If $m_h<10$ GeV, light Higgs cannot decay to $b$-quark pair but can decay to $\tau$-leptons.
In this case, the constraint from the associative production is practically not relevant
because Higgs constraints from $2b2\tau$ final state is much weaker than those from $4b$ final state \cite{LHWG-Note:2005-01}.
On the other hand, the constraint from radiative $\Upsilon$ decay, 
$\Upsilon\to h\gamma$ is on the rise as well as Higgsstrahlung constraint $\sin(\alpha-\beta) \approx 0$. 
Very light scalar particles can contribute the radiative decay of $\Upsilon$-meson \cite{Wilczek:1977zn}.
Including QCD correction \cite{Vysotsky:1980cz,Nason:1986tr,McKeen:2008gd},
the radiative decay of the $\Upsilon$ to the light Higgs is given by
\begin{equation}
\frac{\Gamma(\Upsilon\to h\gamma)}{\Gamma(\Upsilon\to e^+e^-)}
=\frac{m_b^2G_F}{\sqrt{2}\pi\alpha_{\text{em}}}\biggl(\frac{\sin\alpha}{\cos\beta}\biggr)^2
\biggl(1-\frac{m_h^2}{m_{\Upsilon}^2}\biggr)
\biggl[1-\biggl(\frac{\alpha_sC_F}{\pi}\biggr)a_H(z)\biggr] ,
\end{equation}
where $m_b$ is the $b$-quark mass, $G_F$ is the Fermi constant, $\alpha_{\text{em}}$ is the fine structure constant,
$\alpha_s$ is strong coupling, $C_F=4/3$,
$m_{\Upsilon}$ is the $\Upsilon$-meson mass, 
$z=1-{m_h^2}/{m_{\Upsilon}^2}$ and $a_H(z)$ is a QCD correction function which is defined in \cite{Vysotsky:1980cz,Nason:1986tr,McKeen:2008gd}.
Using the fact that ${\cal B}(\Upsilon\to e^+e^-)\simeq0.025$, $\alpha\simeq\beta$ and $a_H(z)\to(4/3)\pi z^{-1/2}+1$ as $z\to0$, we obtain
\begin{equation}
{\cal B}(\Upsilon\to h\gamma)\approx 1.59\times10^{-4} \times z(0.928-0.302z^{-1/2})\tan^2\beta .
\end{equation}
Taking the strongest bound, ${\cal B}(\Upsilon\to h\gamma)<10^{-5}$ \cite{:2008hs,Aubert:2009cka},
we obtain a conservative Higgs mass bound as
\begin{equation}
m_h^2>m_{\Upsilon}^2\biggl[0.894-0.0150\biggl(\frac{\tan\beta}{3}\biggr)^{-2}\biggr]~.
\label{bound:upsilon}
\end{equation}
Hence, we obtain $m_h\gsim8.9$ GeV for $\tan \beta=3$ to evade the radiative $\Upsilon$ decay constraints.

Under such considerations, a neutralino-nucleon scattering cross section 
is obtained in the case of $\sin^2(\alpha-\beta)<0.01$, {\it i.e.}, $\alpha\simeq\beta$, from the eqs. (\ref{feyn:hNN}) and (\ref{feyn:hdd}) such that
\begin{equation}
\begin{split}
\sigma &\simeq \frac{m_N^2 m_r^2}{\pi}\frac{f_{T_s}^2}{2v^2}\frac{g_1^2N_{13}^2N_{11}^2}{m_h^4}\tan^2\beta\\
&\simeq4.7\times10^{-40}\text{ cm}^2\times\biggl(\frac{N_{13}}{0.3}\biggr)^2
\biggl(\frac{\tan\beta}{3}\biggr)^2\biggl(\frac{10\text{ GeV}}{m_h}\biggr)^4,
\end{split}\label{sigma:h}
\end{equation}
for $\tan \beta \sim \mathcal{O}(1)$, as we have expected.

Such a light Higgs scenario explained so far is not easily obtained in ordinary MSSM parameter space.
Instead, we consider BMSSM \cite{Dine:2007xi,Kim:2009sy,Bae:2010cd} where
light Higgs scenarios can be realized \cite{Bae:2010cd}.
The $SU(2)$ doublet CP-even Higgs mass matrix in the basis of $(H_d^0, H_u^0)$ is given by
\begin{equation}
\begin{pmatrix}
M_Z^2c^2_{\beta}+m_A^2s^2_{\beta}-4v^2\epsilon_1s_{2\beta}+4v^2\epsilon_2s^2_{\beta}&
-(M_Z^2+m_A^2)s_{\beta}c_{\beta}-4v^2\epsilon_1\\
-(M_Z^2+m_A^2)s_{\beta}c_{\beta}-4v^2\epsilon_1 &
M_Z^2s^2_{\beta}+m_A^2c^2_{\beta}-4v^2\epsilon_1s_{2\beta}+4v^2\epsilon_2c^2_{\beta}
\end{pmatrix}\label{Higgs:mass}
\end{equation}
where $M_Z$ is the mass of the $Z$ boson, $m_A$ is the mass of the CP-odd Higgs,  $s_{\beta}$($c_{\beta}$) is $\sin\beta$($\cos\beta$), 
and $\epsilon_{1,2}$ are BMSSM parameters defined by \cite{Dine:2007xi}. Here, $\epsilon_{1,2}$ are assumed to be real for simplicity. 
The physical CP-even Higgs bosons are obtained as the mass eigenstates of the above mass matrix,
\begin{equation}
\begin{pmatrix}
H\\h
\end{pmatrix}
=
\begin{pmatrix}
\cos\alpha & \sin\alpha \\ -\sin\alpha & \cos\alpha
\end{pmatrix}
\begin{pmatrix}
H_d^0 \\ H_u^0
\end{pmatrix} \label{eq:pmatrix}
\end{equation}
where $m_h<m_H$ and $-\pi/2 \le \alpha \le \pi/2$ in contrast to the MSSM case.
In the ordinary MSSM case where $\epsilon_{1,2}=0$ and $\tan\beta\lsim 5$, the mass of the CP-odd Higgs $m_A$ cannot be larger than $M_Z$ since $h$ must be light enough to obtain the large WIMP direct detection cross section such as (\ref{sigma:h}). If $m_A\ll M_Z$, light Higgs $h$
is mostly down-type and $\alpha\simeq\pi/2$ so that the constraint from the Higgsstrahlung process $Z^*\to hZ$ can be evaded. But the associative production $e^+e^-\to hA$ still constrains such a case. As previously discussed, it seems possible to avoid this constraint if Higgs bosons decay to neutralinos, however, there still remain other obstacles. If $m_h+m_A<M_Z$, the invisible decay width of $Z$-boson must be considered, hence, such parameter region cannot be viable. 
On the other hand, if $m_A\lsim M_Z$, Higgs mixing is maximized so that $m_h\ll M_Z$ and
$m_H$ is larger than LEP search bound 114 GeV.
In this case, however, the Higgsstrahlung constraint for light Higgs can not be avoided since $\alpha\sim -\pi/4$. Therefore, it seems that very light CP-even Higgs scenario is not realized in the MSSM context.

Considering the BMSSM, instead, the analysis on CP-even Higgs mass is quite different due to the existence of nonzero $\epsilon_{1,2}$.
Observing mass matrix (\ref{Higgs:mass}), the negative $\epsilon_2$ correction effectively reduces $m_A^2$ in the MSSM to $m_A^2+4v^2\epsilon_2$ so that very light $h$ scenario can be realized   
without introducing light CP-odd Higgs. In addition, negative $\epsilon_1$ correction can reduce the off-diagonal Higgs mixing. Moreover, when $4 v^2 |\epsilon_1| \gtrsim (M_Z^2 + m_A^2) s_{\beta} c_{\beta}$, we achieve $\alpha \lsim \pi/2$ so that the condition $\sin^2 (\beta - \alpha) < 0.01$ is satisfied. Therefore, it seems plausible to obtain a light CP-even Higgs scenario in the context of BMSSM.
However, there is a drawback in this scenario.
Negative $\epsilon_2$ can make the Higgs potential unstable along the $D$-flat direction since the quartic term of Higgs potential becomes negative.
In such situation, the electroweak vacuum can become metastable rather than absolute minimum \cite{Blum:2009na}.
We will discuss this issue in Sec. \ref{sec:stability}.

\section{Numerical results in BMSSM parameter space}
\label{sec:lightBMSSM}

In the BMSSM, the $\epsilon_1$ correction is also included in the neutralino sector \cite{Dine:2007xi,Berg:2009mq}.
The coupling $g_{\phi\chi\chi}$ is modified by the $\epsilon_1$ term such that \cite{Berg:2009mq}
\begin{eqnarray}
\delta g_{h\chi\chi} &=& -\frac{2\epsilon_1}{\mu}\big(v\sqrt{2}\cos\beta\cos\alpha (N_{14})^2
+v\sqrt{2}\sin\beta\sin\alpha (N_{13})^2\nonumber\\
&&+2\sqrt{2}v\sin(\alpha+\beta)N_{13}N_{14}\big),\label{feyn:e1_hNN}\\
\delta g_{A\chi\chi} &=& -\frac{2\epsilon_1}{\mu}\big(iv\frac{1}{\sqrt{2}}\sin2\beta(N_{14})^2
+iv\frac{1}{\sqrt{2}}\sin2\beta(N_{13})^2+i2\sqrt{2}vN_{13}N_{14}\big),\label{feyn:e1_ANN}
\end{eqnarray}
where $\mu$ is the Higgsino mass parameter in the MSSM superpotential.
In our circumstance, the BMSSM corrections for neutralino-Higgs interaction cancel the MSSM interactions
in parameter space since $\alpha>0$ and $\epsilon_1>0$ in eqs. (\ref{feyn:hNN}), (\ref{feyn:ANN}),
(\ref{feyn:e1_hNN}) and (\ref{feyn:e1_ANN}).
However, since the BMSSM corrections are much smaller than the MSSM ones, these do not spoil the aforementioned advantages.
\begin{figure}
\subfigure[\ $\tan\beta=3$, $m_A=90$ GeV]{
\includegraphics[width=7.5cm]{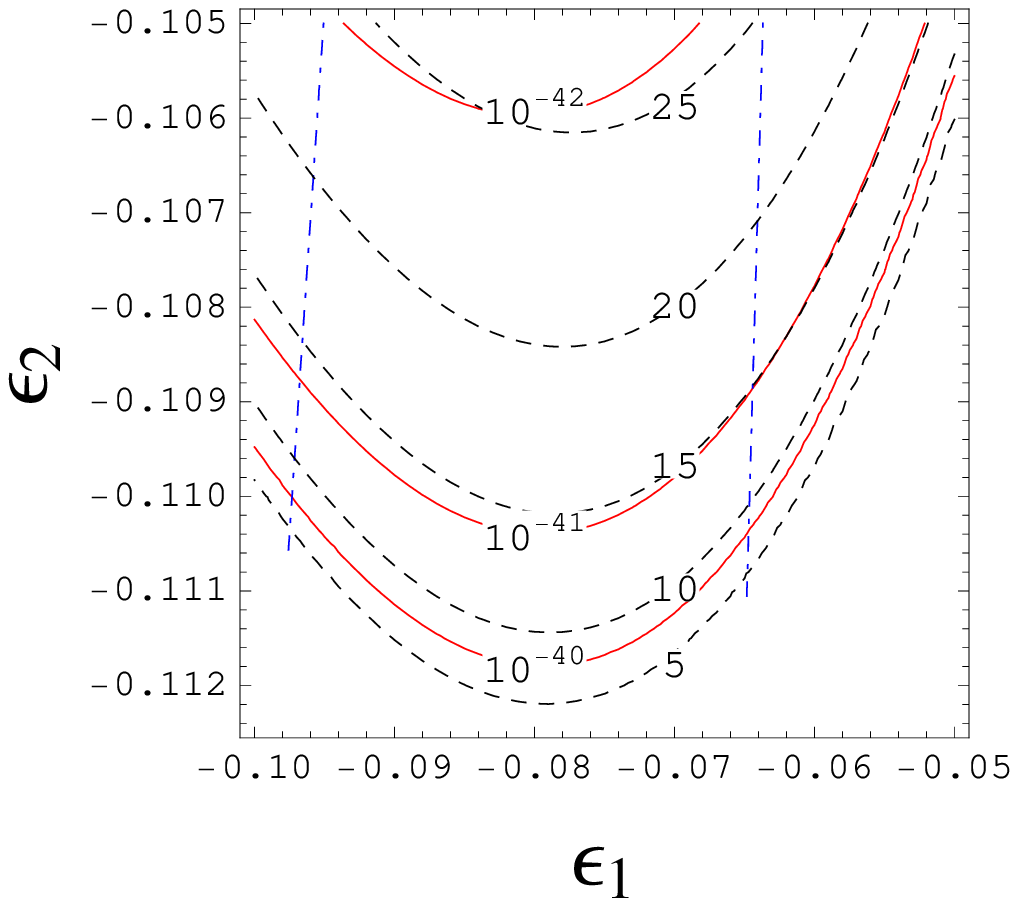}}
\quad
\subfigure[\ $\tan\beta=5$, $m_A=90$ GeV]{
\includegraphics[width=7.5cm]{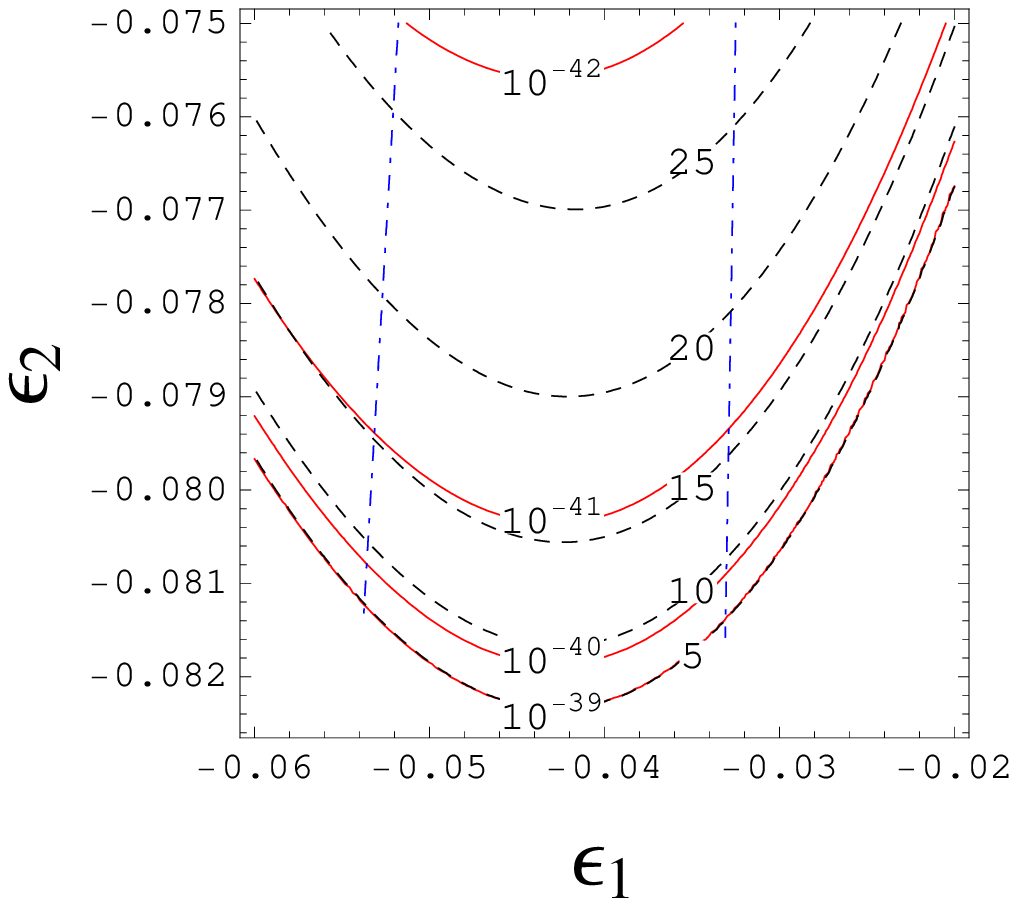}}
\caption{Numerical results of neutralino-nucleon scattering cross section in the 
$\epsilon_1-\epsilon_2$ plane. Solid(red) curves stand for scattering cross section,
$\sigma=10^{-40}$-$10^{-42}$ cm$^2$ for left panel and $\sigma=10^{-39}$-$10^{-42}$ cm$^2$ for right panel
from the bottom to the top.
Dashed(black) curves stand for light Higgs mass, $m_h=5$-$25$ GeV 
from the bottom to the top. Dot-dashed(blue) lines stand for $\sin^2(\alpha-\beta)=0.01$,
the region between the lines is safe from the Higgsstrahlung constraint.
Neutralino mass is $4$ GeV$\lsim m_{\chi} \lsim7$ GeV depending on parameters.}
\label{fig:result}
\end{figure}
Numerical results are given in Fig. \ref{fig:result}.
From the figures, $-0.10\lesssim\epsilon_1\lsim -0.06$, $\epsilon_2\sim-0.11$ for $\tan\beta=3$
and $-0.06 \lesssim\epsilon_1\lesssim -0.03$, $\epsilon_2\sim-0.08$ for $\tan\beta=5$
give the desired scattering cross section for CoGeNT results, simultaneously  satisfying the LEP Higgsstrahlung constraint.

In the case of $m_h>10$ GeV, however, this parameter space is not allowed due to the large value of $R_{hA}\sim0.3>0.2$. We, hence, need a lower value of $\tan \beta$ since the branching ratio of $h$ to $\chi \chi$ becomes larger for smaller $\tan \beta$. Comparing Fig. \ref{fig:result}(a) and Fig. \ref{fig:result}(b), however, larger values of $|\epsilon_{1,2}|$ are needed to obtain the light Higgs spectrum for smaller $\tan \beta$. Then, we need much fine tuning to obtain such light Higgs so that it is more natural to consider only the case $m_h<10$ GeV. In this range of mass, the LEP constraint from the associative production is avoided. Here, we must tune the light Higgs mass more or less in order to satisfy the conservative bound of radiative $\Upsilon$ decay such as eq. (\ref{bound:upsilon}).
For the larger $\tan\beta$ case, we also have desired the neutralino-nucleon cross section satisfying all
experimental constraints for $\sin^2(\alpha-\beta)<0.01$ and $9$ GeV$\lsim m_h\lsim10$ GeV. Consequently, the light Higgs scenario to explain the light neutralino of $4$ GeV $\lsim m_{\chi} \lsim$ $7$ GeV and $\sigma_{\text{SI}} \sim 10^{-40}\text{ cm}^2$ is most naturally realized when $\sin^2(\alpha-\beta)<0.01$ and $9$ GeV$\lsim m_h\lsim10$ GeV.

\section{Stability of the BMSSM scalar potential}
\label{sec:stability} 

The analysis of the light BMSSM neutralino so far is based on the assumption that we have already chosen the stable vacuum of the scalar potential. This issue is, however, nontrivial since the stability of the scalar potential highly depends on the coefficients of the leading nonrenormalizable operators \cite{Blum:2009na,Bernal:2009hd}. Therefore it is required to discuss whether our light neutralino with light Higgs can be realized simultaneously respecting the stability of the potential. 

According to the criterion of \cite{Blum:2009na}, $\epsilon_2\pm\epsilon_1/4>0$ in order to have positive quartic term along the $D$-flat directions.
However, we prefer large negative $\epsilon_2$, as we have seen in the previous section, to make the CP-even Higgs light 
so that the electroweak vacuum becomes meta stable.
In order to cure such a drawback, we can consider a more general Higgs scalar potential \cite{Bae:2010cd},
\begin{equation}
\begin{split}
\delta V =& 2\epsilon_1H_uH_d(H_u^{\dagger}H_u+H_d^{\dagger}H_d)+\text{h.c.}\\
&+\epsilon_2(H_uH_d)^2+\text{h.c.}\\
&+\epsilon_3(H_u^{\dagger}H_u)^2+\epsilon_4H_uH_d(H_uH_d)^{\dagger}.
\end{split}
\end{equation}
The $\epsilon_3$ and $\epsilon_4$ terms can be considered effective dimension six operators of MSSM
so that we can easily introduce such interactions \cite{Antoniadis:2009rn}.
Further, such dimension six operators can have sizeable effects according to the parameter space \cite{Antoniadis:2009rn}.
In this work, we do not study microscopic models which can make such effective operators.
Instead we just parametrize such terms and analyze the parameter space that can make CP-even Higgs as light as 10 GeV.
The $\epsilon_3$ term has the same effect as a top-stop loop contribution to MSSM Higgs mass, so we neglect it.

The effective Higgs potential along the MSSM D-flat directions at tree level is given as 
\begin{eqnarray}
V^{\text{D-flat}} (\phi) = \frac12 \left( m_1^2 + m_2^2 \mp 2 m_{12}^2 \right) \phi^2 + 2 \left( \frac{\epsilon_4}{8}+\frac{\epsilon_2}{4} \mp \epsilon_1 \right) \phi^4 + \left| \frac{\epsilon_1}{\mu} \right|^2 \phi^6 ,
\end{eqnarray}
ignoring $\epsilon_3$, where $\pm v_u = v_d \equiv \phi/\sqrt{2}$, $m_1^2 = m_{H_d}^2 +|\mu|^2$, $m_2^2 = m_{H_u}^2$, $m_{12}^2 = B\mu$. If the quartic coupling of the above equation is positive, there is no additional vacuum away from the electroweak scale. Therefore the most conservative bound is given as
\begin{eqnarray}
\frac{\epsilon_4}{8}+\frac{\epsilon_2}{4} \mp \epsilon_1 > 0 .
\label{eq:stable}
\end{eqnarray}

Including the $\epsilon_4$ correction, the Higgs mass matrix becomes
\begin{equation}
\begin{pmatrix}
M_Z^2c^2_{\beta}+m_A^2s^2_{\beta}-4v^2\epsilon_1s_{2\beta}+4v^2(\epsilon_2 + \frac14 \epsilon_4)s^2_{\beta}&
-(M_Z^2+m_A^2)s_{\beta}c_{\beta}-4v^2\epsilon_1 + 3\epsilon_4 v^2 c_{\beta}s_{\beta}\\
-(M_Z^2+m_A^2)s_{\beta}c_{\beta}-4v^2\epsilon_1 + 3\epsilon_4 v^2 c_{\beta}s_{\beta}&
M_Z^2s^2_{\beta}+m_A^2c^2_{\beta}-4v^2\epsilon_1s_{2\beta}+4v^2(\epsilon_2 + \frac14 \epsilon_4)c^2_{\beta}
\end{pmatrix}\label{eq:stablehiggs}
\end{equation}

\begin{figure}
\subfigure[\ $\tan\beta=10$, $m_A=70$ GeV, $\epsilon_4=0.2$]{
\includegraphics[width=7.5cm]{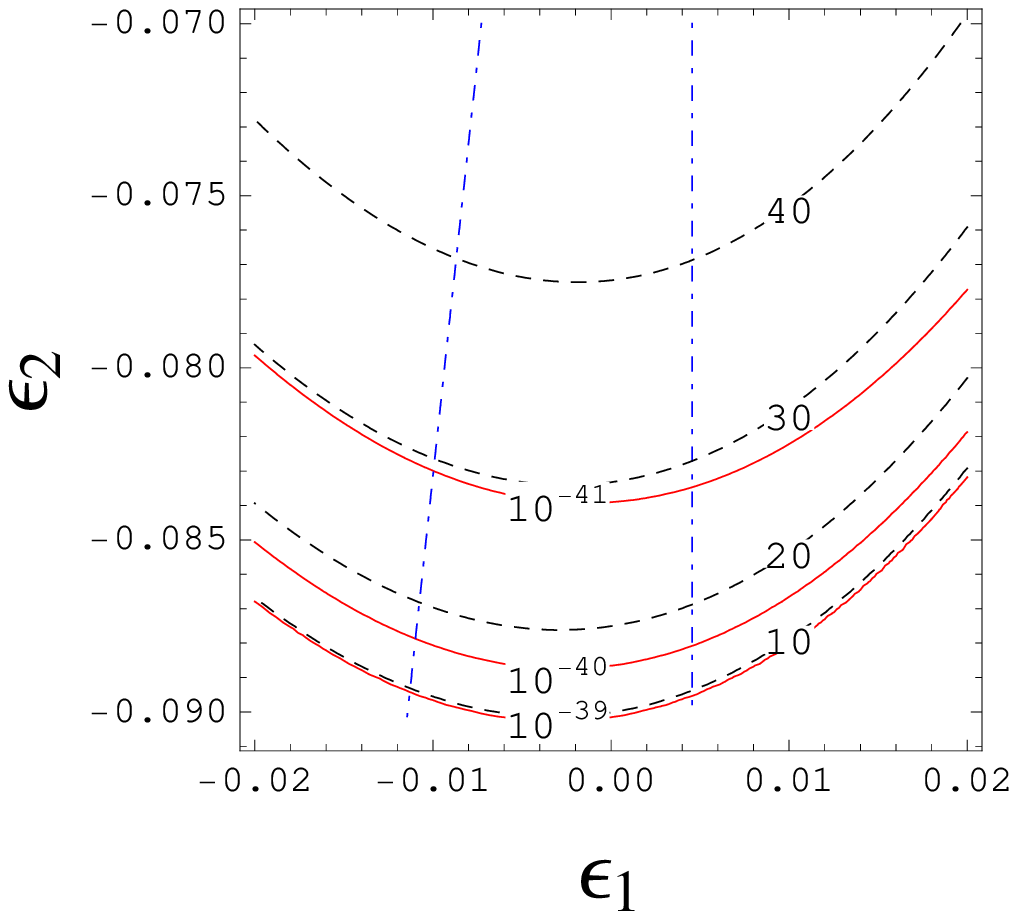}}
\quad
\subfigure[\ $\tan\beta=10$, $m_A=90$ GeV, $\epsilon_4=0.3$]{
\includegraphics[width=7.5cm]{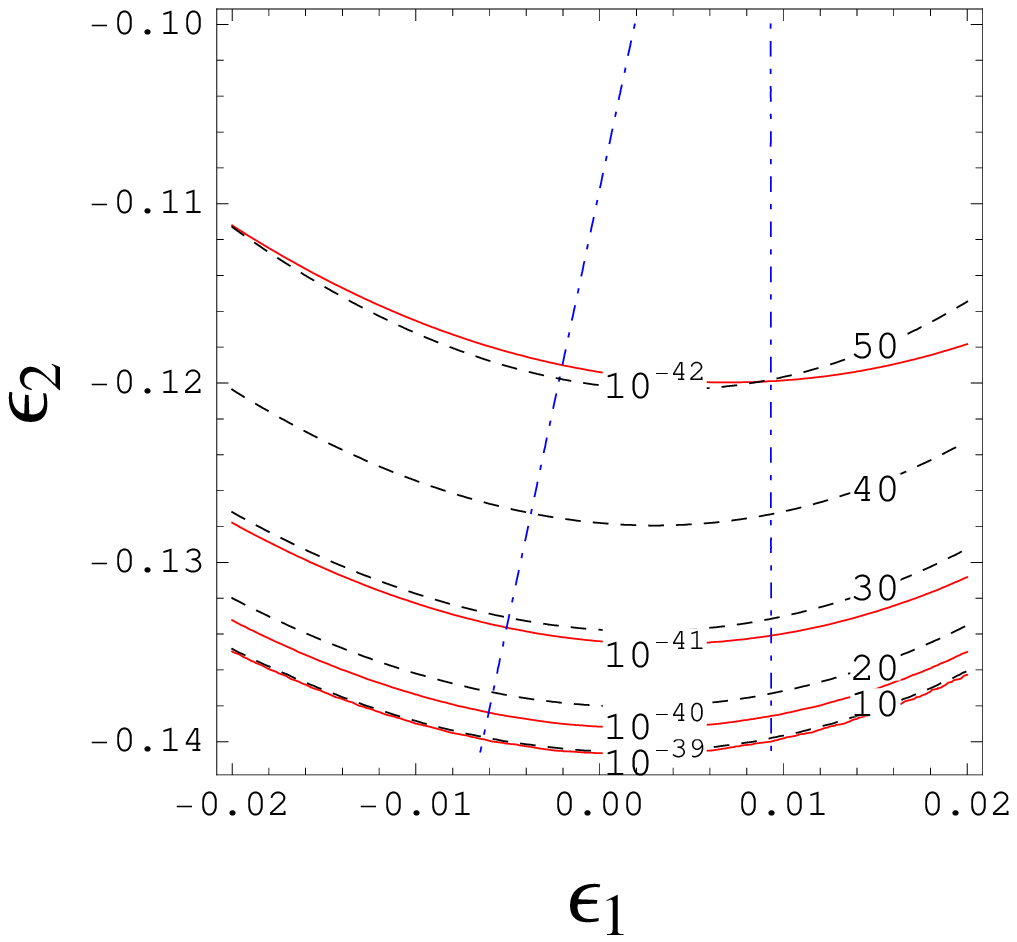}}
\caption{The case for $\tan\beta=10$. The curves are defined the same as for Fig. \ref{fig:result}.}
\label{fig:stability}
\end{figure}

To obtain the similar mass spectrum as the previous parameter space expected in the previous section, we obtain two large values of $\epsilon_{2,4}$ to satisfy eq.(\ref{eq:stable}), which do not seem to be under control in the BMSSM. If we take the larger value of $\tan\beta$, then such corrections can be lowered. As an example, we analyzed the case for $\tan\beta=10$ in Fig. \ref{fig:stability}. Since the mass of our light Higgs is around or smaller than 10 GeV, the branching ratio of $h \to \chi \chi$ does not need to be dominant so that $\tan\beta$ can be higher than 3. The value of $\tan\beta > 10$ may enhance the cross section $\sigma_{\small \text{SI}}$ too much but $\tan\beta \sim 10$ is a proper value to obtain the $\sigma_{\small \text{SI}} \sim 10^{-40} \text{cm}^2$ once we lower the value of $N_{13}$.

\section{Relic Abundance}
\label{sec:relic}

In this section, we compute the right relic abundance of the neutralino. Since the neutralino is very light, $m_{\chi}\lsim7$ GeV, they annihilate only to light fermions at the freeze-out. In addition, the neutralino is much lighter than the CP-odd Higgs $A$, squarks, and $Z$-boson so that the dominant annihilation process is mediated by the CP-even Higgs $h$ which is a P-wave process. Furthermore, small $\tan\beta \sim 3$ constrains the interaction of $h$ to the SM fermions. Therefore, one might worry that the neutralino will overclose the universe. There is one way out, however. Since the mass of the light CP-even Higgs is highly constrained, $9 \mbox{ GeV} \lsim m_h \lsim 10\mbox{ GeV}$, the resonant annihilation of the light WIMPs to the SM fermions through the $s$-channel process can dominate the annihilation process and reduce the relic abundance at the freeze-out although the process is P-wave suppressed. 

In our case, the thermal average of WIMP annihilation cross section at the freeze-out is not obtained by the nonrelativistic approximation called standard calculation in \cite{LSP1,Lee:1977ua}. Considering the resonant annihilation of WIMPs through the $s$-channel process, the thermal average of $\sigma v$ is affected by the values of $v$ not close to the average value. Therefore, the following full calculation is needed as described in \cite{Griest:1990kh}.
\begin{eqnarray}
\Omega h^2 &=& \frac{1.07 \times 10^9 \text{ GeV}^{-1}}{J(x_f) g_{\ast}^{1/2} m_{\text{Pl}}} ,  \label{eq:numOh}
\end{eqnarray} 
where 
\begin{eqnarray}
J(x_f) &=& \int_{x_f}^{\infty} \frac{\langle \sigma v \rangle}{x^2} dx 
= \int_{0}^{\infty} dv \frac{v^2 (\sigma v)}{\sqrt{4\pi}} \int_{x_f}^{\infty} dx \ x^{-1/2} e^{-x v^2/4} , 
\end{eqnarray}
$m_{\text{Pl}}=1.22\times10^{19}$ GeV is the Planck mass, and 
$g_{\ast}=62.625$ is the effective number of relativistic degrees of freedom at the freeze-out temperature $T_f \sim 0.3$ GeV. The inverse freeze-out temperature $x_f=m_{\chi}/T_f$ is determined by the iterative equation 
\begin{eqnarray}
x_f = \log \left( \frac{m_{\chi}}{2\pi^3} 
                  \sqrt{\frac{45 m_{\text{Pl}}^2}{2 g_{\ast} x_f}} 
                  \langle \sigma v \rangle \right) ,
\end{eqnarray}
where the assumption that $x_f \sim 20$ obtains
\begin{eqnarray}
x_f & \approx & \log \left(0.0765 \langle \sigma v \rangle m_{\text{Pl}} m_{\chi} / g_{\ast}^{1/2} \right) - \frac12 \log x_f \nonumber \\
& = & \log \left(0.0765 \langle \sigma v \rangle m_{\text{Pl}} m_{\chi} / g_{\ast}^{1/2} \right) - \frac12 \left( \log 20 + \frac{x_f - 20}{20} - \frac{(x_f -20)^2}{800} + \cdots \right) \nonumber ,
\end{eqnarray}
where we use the series expansion at $x_f=20$ in the second line.
Hence, we approximately obtain
\begin{eqnarray}
x_f \approx 19.6 + \log \frac{\langle \sigma v \rangle}{10^{-9} \text{ GeV}^{-2}} + \log \frac{m_{\chi}}{10 \text{GeV}} - \frac12 \log \frac{100}{g_{\ast}} .
\end{eqnarray}

\begin{figure}
\begin{center}
\includegraphics[width=14cm]{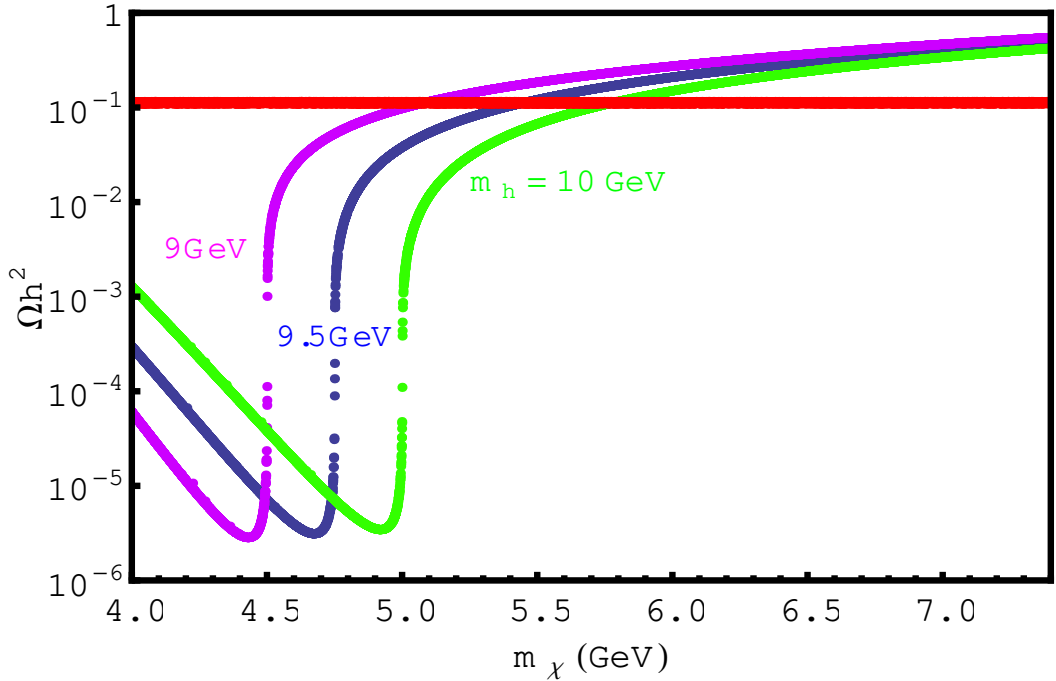}
\end{center}
\caption{$\Omega h^2$ to $m_{\chi}$ with fixed $\tan \beta =3$ and $N_{13} = 0.3$ values. The allowed mass of the light CP-even Higgs is $9$ GeV $\lsim m_h \lsim$ $10$ GeV. The magenta, blue, green lines denote the case $m_h = 9$ GeV, 9.5 GeV, 10 GeV respectively. The parameters are given to avoid the LEP or $\Upsilon$ decay constraints. The red region denotes the observed relic abundance satisfying (\ref{eq:DMrelic}). Therefore, we conclude that the physically consistent mass of our light WIMP is determined as $5$ GeV$ \lsim m_{\chi} \lsim$ $6$ GeV here.}
\label{fig:relic}
\end{figure}

The annihilation cross section of the neutralinos to $b$-quark pairs dominantly determines the present relic abundance within 10 \% accuracy. The rest of the contributions are from the annihilations to the lighter fermions such as $\tau$-lepton and $c$-quark pairs. Therefore, we obtain the cross section such that 
\begin{eqnarray}
\sigma v & = & \frac{3}{16 \pi s} (g_{h \chi \chi} + \delta g_{h \chi \chi})^2 g_{h b b}^2 \frac{s^2}{(s-m_h^2)^2 + \Gamma_h^2 m_h^2} \left( 1 - \frac{4 m_b^2}{s}\right)^{3/2} \left( 1 - \frac{4 m_{\chi}^2}{s}\right) , \label{eq:sigv}
\end{eqnarray}
where the center of mass energy $\sqrt{s} = 2m_{\chi}/\sqrt{1-v^2/4}$ and $\Gamma_h$ is the total decay width of the CP-even Higgs $h$ such that
\begin{eqnarray}
\Gamma_h &=& \frac{m_h}{16 \pi \cdot 4} (g_{h \chi \chi} + \delta g_{h \chi \chi})^2
\left( 1 - \frac{4 m_{\chi}^2}{m_h^2}\right)^{3/2} + \frac{3 m_h }{8\pi} g_{hbb}^2 \left( 1 - \frac{4 m_b^2}{m_h^2}\right)^{3/2} ,
\end{eqnarray}
where the contributions from lighter fermion final states are ignored.

We show the role of resonant annihilation to obtain the observed relic density of (\ref{eq:DMrelic}) in Fig.\ref{fig:relic}. The parameter space is chosen to satisfy the LEP and $\Upsilon$ decay constraints. In this figure, we fix $\tan\beta=3$, $N_{13} = 0.3$ values. The allowed mass of the light CP-even Higgs is $9$ GeV $\lsim m_h \lsim$ $10$ GeV. The magenta, blue, green lines denote the case $m_h = 9$ GeV, 9.5 GeV, 10 GeV, respectively. The parameters are given to avoid the LEP or $\Upsilon$ decay constraints. The red region denotes the observed relic abundance satisfying (\ref{eq:DMrelic}). It is clear that the expected relic abundance decreases around the resonance region so that the mass of the light neutralino is determined within $5$ GeV$ \lsim m_{\chi} \lsim$ $6$ GeV. The physically consistent range of $m_{\chi}$ is similar for other possible parameter choice since the right relic abundance will be obtained within the the resonant annihilation process.   

The QCD phase transition occurs around $T\lsim0.3$ GeV$\sim T_f$ in our favorite parameter region, which might be a dangerous threat to our calculation of (\ref{eq:sigv}) since we have to consider the hadronization of quarks. Fortunately, the recent result in lattice QCD method \cite{Karsch:2000kv} informs us that the critical temperature of QCD phase transition is as low as $0.15$ GeV when we consider the number of flavor is three.   

Now, it is fair to explain the situations after the first version of this paper was submitted to the e-print arXiv. It is claimed that the scintillation efficiency $\mathcal{L}_{\text{eff}}$ in the XENON detectors are not so much suppressed above 3.4 keV nuclear recoil energy so that even XENON10 can provide the strong exclusion bound to the WIMP heavier than 6 GeV \cite{Savage:2010tg}. In addition, there is another interesting approach avoiding the discussions on the $\mathcal{L}_{\text{eff}}$ by focussing on the S2 only signal in XENON10 \cite{Sorensen:2010hq}. There are two PMTs in XENON experiments, called S1 and S2. The S1 measures the scintillation inside the detector, while the S2 measures the ionization signal by which the discrimination of the background signals is well performed. Since the $\mathcal{L}_{\text{eff}}$ is only related to the scintillation, the S2 only analysis is a very fancy idea to probe the ``exclusion" bound of the light WIMPs in contrast to the discovery region. 
None of the proposals, however, rules out the WIMP scattering signals whose mass is $\lesssim 5 - 6$ GeV and the elastic scattering cross section of $10^{-40} \text{cm}^2$, which can be expected by considering proper background contaminations in the CoGeNT results. In this sense, our neutralino of mass 5 to 6 GeV is quite promising in the search for viable light WIMPs.

\section{Conclusions}
\label{sec:conclusions}

The light WIMPs with large cross section are focused on due to the recent results of direct detection experiments such as CoGeNT and DAMA. Since SUSY is one of the most promising candidates of the theory beyond the SM, light neutralino should be carefully studied. In this sense, we have shown that such a light neutralino can be obtained from the light CP-even Higgs scenario. In the context of the MSSM, however, the existence of such a light Higgs is highly constrained by the LEP experiments. Instead, we looked for a possibility of explaining $\sigma_{\text{SI}} \sim 10^{-40} \text{ cm}^2$, $m_{\chi} \sim$ 4 to 7 GeV dark matter within the framework of the BMSSM (MSSM field contents at and below the weak scale)
and found that $m_h \sim$ 9 to 10 GeV can provide the required $\sigma_{\text{SI}}$. If we require the model to explain the right relic abundance, $m_{\chi}$ is predicted to be in between 5 to 6 GeV
depending on the light CP even Higgs mass from 9 to 10 GeV.

Combined with the improvement of the detector parameters such as $\mathcal{L}_{\text{eff}}$ at low energy, more data in the current or future direct detection experiments will eventually reveal whether the light dark matter scenario inspired by CoGeNT and DAMA/LIBRA is realized in nature or not.
We pointed out that it is possible to come up with a model in supersymmetric theories without introducing an additional light degree of freedom in addition to the MSSM at and below the weak scale.

\begin{acknowledgements}
This work is supported by KRF-2008-313-C00162 and NRF with CQUEST 2005-0049409 (KJB and HDK). 
\end{acknowledgements}


\begin{thebibliography}{00}

\bibitem{Zwicky:1933gu}
  F.~Zwicky,
  Helv.\ Phys.\ Acta {\bf 6}, 110-127 (1933); 
  F.~Zwicky,
  Astrophys.\ J.\  {\bf 86}, 217-246 (1937). 

\bibitem{rotation} 
  Y.~Sofue and V.~Rubin,
  Ann.\ Rev.\ Astron.\ Astrophys.\  {\bf 39}, 137-174 (2001) 
  [astro-ph/0010594];
  A.~Borriello and P.~Salucci,
  Mon.\ Not.\ Roy.\ Astron.\ Soc.\  {\bf 323}, 285 (2001) 
  [astro-ph/0001082].

\bibitem{Bullet} Bullet Cluster Collaboration (M. Bradac for the collaboration), "Shedding light on dark matter: Seeing the invisible with the Bullet Cluster 1E0657-56", 
*Heidelberg 2007, Dark matter in astroparticle and particle physics* 254-259 ;
  D.~Clowe, M.~Bradac and A.~H.~Gonzalez {\it et al.},
  Astrophys.\ J.\  {\bf 648}, L109-L113 (2006) 
  [astro-ph/0608407].

\bibitem{WMAP} 
  E.~Komatsu, K.~M.~Smith and J.~Dunkley {\it et al.},
  [arXiv:1001.4538 [astro-ph.CO]].

\bibitem{LSP1} 
  G.~Jungman, M.~Kamionkowski and K.~Griest,
  Phys.\ Rept.\  {\bf 267}, 195-373 (1996) 
  [hep-ph/9506380].

\bibitem{LSP2}
  G.~Bertone, D.~Hooper and J.~Silk,
  Phys.\ Rept.\  {\bf 405}, 279-390 (2005) 
  [hep-ph/0404175];
  J.~R.~Ellis, K.~A.~Olive and Y.~Santoso {\it et al.},
  Phys.\ Lett.\  {\bf B565}, 176-182 (2003) 
  [hep-ph/0303043].

\bibitem{LKKP} 
  H.~-C.~Cheng, J.~L.~Feng and K.~T.~Matchev,
  Phys.\ Rev.\ Lett.\  {\bf 89}, 211301 (2002) 
  [hep-ph/0207125];
  G.~Servant and T.~M.~P.~Tait,
  Nucl.\ Phys.\  {\bf B650}, 391-419 (2003) 
  [hep-ph/0206071];
\%bibitem{Burnell:2005hm}
  F.~Burnell and G.~D.~Kribs,
  Phys.\ Rev.\  {\bf D73}, 015001 (2006) 
  [hep-ph/0509118];
  K.~Kong and K.~T.~Matchev,
  JHEP {\bf 0601}, 038 (2006) 
  [hep-ph/0509119].

\bibitem{Todd}  
  H.~-C.~Cheng and I.~Low,
  JHEP {\bf 0309}, 051 (2003) 
  [hep-ph/0308199]; 
  H.~-C.~Cheng and I.~Low,
  JHEP {\bf 0408}, 061 (2004) 
  [hep-ph/0405243].

\bibitem{SDM} 
  Y.~G.~Kim, K.~Y.~Lee and S.~Shin,
  JHEP {\bf 0805}, 100 (2008) 
  [arXiv:0803.2932 [hep-ph]];
  K.~Y.~Lee, Y.~G.~Kim and S.~Shin,
  [arXiv:0809.2745 [hep-ph]]; 
  Y.~G.~Kim and K.~Y.~Lee,
  Phys.\ Rev.\  {\bf D75}, 115012 (2007) 
  [hep-ph/0611069];
V. Silveira and A. Zee, Phys. Lett. {\bf B161}, 136 (1985);
  C.~P.~Burgess, M.~Pospelov and T.~ter Veldhuis,
  Nucl.\ Phys.\  {\bf B619}, 709-728 (2001) 
  [hep-ph/0011335];
  H.~Davoudiasl, R.~Kitano and T.~Li {\it et al.},
  Phys.\ Lett.\  {\bf B609}, 117-123 (2005) 
  [hep-ph/0405097].

\bibitem{Lee:1977ua}
  B.~W.~Lee and S.~Weinberg,
  Phys.\ Rev.\ Lett.\  {\bf 39}, 165-168 (1977).

\bibitem{cogent} 
  C.~E.~Aalseth {\it et al.} [ CoGeNT Collaboration ],
  [arXiv:1002.4703 [astro-ph.CO]].

\bibitem{dama} 
  R.~Bernabei {\it et al.} [ DAMA Collaboration ],
  Eur.\ Phys.\ J.\  {\bf C56}, 333-355 (2008) 
  [arXiv:0804.2741 [astro-ph]].

\bibitem{channeling} 
  R.~Bernabei, P.~Belli and F.~Montecchia {\it et al.},
  Eur.\ Phys.\ J.\  {\bf C53}, 205-213 (2008) 
  [arXiv:0710.0288 [astro-ph]].

\bibitem{SFDMdama}
  Y.~G.~Kim and S.~Shin,
  JHEP {\bf 0905}, 036 (2009) 
  [arXiv:0901.2609 [hep-ph]].
  
\bibitem{andreas} 
  S.~Andreas, T.~Hambye and M.~H.~G.~Tytgat,
  JCAP {\bf 0810}, 034 (2008) 
  [arXiv:0808.0255 [hep-ph]].

\bibitem{wimpless} 
  J.~L.~Feng and J.~Kumar,
  Phys.\ Rev.\ Lett.\  {\bf 101}, 231301 (2008) 
  [arXiv:0803.4196 [hep-ph]]; 
  J.~L.~Feng, J.~Kumar and L.~E.~Strigari,
  Phys.\ Lett.\  {\bf B670}, 37-40 (2008) 
  [arXiv:0806.3746 [hep-ph]].

\bibitem{bottino}
  A.~Bottino, F.~Donato, N.~Fornengo and S.~Scopel,
  Phys.\ Rev.\  {\bf D78}, 083520 (2008) 
  [arXiv:0806.4099 [hep-ph]].

\bibitem{cerdeno} 
  D.~G.~Cerdeno, C.~Munoz and O.~Seto,
  Phys.\ Rev.\  {\bf D79}, 023510 (2009) 
  [arXiv:0807.3029 [hep-ph]].

\bibitem{foot} 
  R.~Foot,
  Phys.\ Rev.\  {\bf D74}, 023514 (2006) 
  [astro-ph/0510705];
  R.~Foot,
  Phys.\ Rev.\  {\bf D78}, 043529 (2008) 
  [arXiv:0804.4518 [hep-ph]].

\bibitem{Bozorgnia:2010xy}
  N.~Bozorgnia, G.~B.~Gelmini and P.~Gondolo,
  [arXiv:1006.3110 [astro-ph.CO]].'
  
\bibitem{lightnmssm}
  J.~F.~Gunion, D.~Hooper, B.~McElrath,
  Phys.\ Rev.\  {\bf D73}, 015011 (2006)  
  [hep-ph/0509024];
  A.~V.~Belikov, J.~F.~Gunion, D.~Hooper {\it et al.},
  [arXiv:1009.0549 [hep-ph]];
  J.~F.~Gunion, A.~V.~~Belikov and D.~Hooper,
  [arXiv:1009.2555 [hep-ph]].

\bibitem{cogentpapers1} 
  A.~L.~Fitzpatrick, D.~Hooper and K.~M.~Zurek,
  [arXiv:1003.0014 [hep-ph]];
  J.~Kopp, T.~Schwetz and J.~Zupan,
  JCAP {\bf 1002}, 014 (2010) 
  [arXiv:0912.4264 [hep-ph]];
  S.~Andreas, C.~Arina, T.~Hambye {\it et al.},
  [arXiv:1003.2595 [hep-ph]];
  R.~Foot,
  [arXiv:1004.1424 [hep-ph]];
  V.~Barger, M.~McCaskey, G.~Shaughnessy,
  [arXiv:1005.3328 [hep-ph]].

\bibitem{cogentpapers2} 
  S.~Chang, J.~Liu, A.~Pierce {\it et al.},
  [arXiv:1004.0697 [hep-ph]].
  
\bibitem{cdmssi} 
  D.~S.~Akerib {\it et al.} [ CDMS Collaboration ],
  Phys.\ Rev.\ Lett.\  {\bf 96}, 011302 (2006) 
  [astro-ph/0509259].

\bibitem{xenon100} 
  E.~Aprile {\it et al.} [ XENON100 Collaboration ],
  [arXiv:1005.0380 [astro-ph.CO]].

\bibitem{xenonreply} 
  T.~X.~Collaboration,
  [arXiv:1005.2615 [astro-ph.CO]].

\bibitem{collar:2010} 
  J.~I.~Collar and D.~N.~McKinsey,
  [arXiv:1005.0838 [astro-ph.CO]];
  J.~I.~Collar and D.~N.~McKinsey,
  [arXiv:1005.3723 [astro-ph.CO]].

\bibitem{Lebedenko:2008gb}
  V.~N.~Lebedenko, H.~M.~Araujo and E.~J.~Barnes {\it et al.},
  Phys.\ Rev.\  {\bf D80}, 052010 (2009) 
  [arXiv:0812.1150 [astro-ph]].

\bibitem{Ellis} 
  J.~R.~Ellis, A.~Ferstl and K.~A.~Olive,
  Phys.\ Lett.\  {\bf B481}, 304-314 (2000) 
  [hep-ph/0001005].

\bibitem{Drees:2004jm}
  M.~Drees, R.~Godbole and P.~Roy,
  Hackensack, USA: World Scientific (2004) 555 p.

\bibitem{Bae:2007pa}
  K.J.~Bae, R.~Dermisek, H.D.~Kim and I.W.~Kim,
  JCAP {\bf 0708}, 014 (2007)
  [arXiv:hep-ph/0702041];
  R.~Dermisek, H.~D.~Kim and I.~W.~Kim,
  JHEP {\bf 0610}, 001 (2006)
  [arXiv:hep-ph/0607169].


\bibitem{Feldman:2010ke}
  D.~Feldman, Z.~Liu and P.~Nath,
  arXiv:1003.0437 [hep-ph].

\bibitem{Kuflik:2010ah}
  E.~Kuflik, A.~Pierce and K.~M.~Zurek,
  arXiv:1003.0682 [hep-ph].

\bibitem{Bae:2010ai}
  K.J.~Bae,
  arXiv:1003.5869 [hep-ph].

\bibitem{Dreiner:2009ic}
  H.~K.~Dreiner, S.~Heinemeyer, O.~Kittel, U.~Langenfeld, A.~M.~Weber and G.~Weiglein,
  Eur.\ Phys.\ J.\  C {\bf 62}, 547 (2009)
  [arXiv:0901.3485 [hep-ph]].

\bibitem{LHWG-Note:2005-01}
  http://lephiggs.web.cern.ch/LEPHIGGS/papers/July2005\_MSSM/index.html

\bibitem{Yaguna:2007vm}
  C.~E.~Yaguna,
  Phys.\ Rev.\  D {\bf 76}, 075017 (2007)
  [arXiv:0708.0248 [hep-ph]].

\bibitem{Wilczek:1977zn}
  F.~Wilczek,
  Phys.\ Rev.\ Lett.\  {\bf 39}, 1304 (1977).

\bibitem{Vysotsky:1980cz}
  M.~I.~Vysotsky,
  Phys.\ Lett.\  {\bf B97}, 159-162 (1980).

\bibitem{Nason:1986tr}
  P.~Nason,
  Phys.\ Lett.\  {\bf B175}, 223 (1986).

\bibitem{McKeen:2008gd}
  D.~McKeen,
  Phys.\ Rev.\  {\bf D79}, 015007 (2009) 
  [arXiv:0809.4787 [hep-ph]].

\bibitem{:2008hs}
  W.~Love {\it et al.} [ CLEO Collaboration ],
  Phys.\ Rev.\ Lett.\  {\bf 101}, 151802 (2008) 
  [arXiv:0807.1427 [hep-ex]].

\bibitem{Aubert:2009cka}
  B.~Aubert {\it et al.}  [BABAR Collaboration],
  Phys.\ Rev.\ Lett.\  {\bf 103}, 181801 (2009)
  [arXiv:0906.2219 [hep-ex]].

\bibitem{Dine:2007xi}
  M.~Dine, N.~Seiberg and S.~Thomas,
  Phys.\ Rev.\  D {\bf 76}, 095004 (2007)
  [arXiv:0707.0005 [hep-ph]].

\bibitem{Kim:2009sy}
  H.D.~Kim and J.H.~Kim,
  JHEP {\bf 0905}, 040 (2009)
  [arXiv:0903.0025 [hep-ph]].

\bibitem{Bae:2010cd}
  K.J.~Bae, R.~Dermisek, D.~Kim, H.D.~Kim and J.H.~Kim,
  arXiv:1001.0623 [hep-ph].

\bibitem{Blum:2009na}
  K.~Blum, C.~Delaunay, Y.~Hochberg,
  Phys.\ Rev.\  {\bf D80}, 075004 (2009) 
  [arXiv:0905.1701 [hep-ph]].

\bibitem{Berg:2009mq}
  M.~Berg, J.~Edsjo, P.~Gondolo, E.~Lundstrom and S.~Sjors,
  JCAP {\bf 0908}, 035 (2009)
  [arXiv:0906.0583 [hep-ph]].

\bibitem{Bernal:2009hd}
  N.~Bernal, K.~Blum, Y.~Nir {\it et al.},
  JHEP {\bf 0908}, 053 (2009) 
  [arXiv:0906.4696 [hep-ph]].

\bibitem{Antoniadis:2009rn}
  I.~Antoniadis, E.~Dudas, D.~M.~Ghilencea {\it et al.},
  Nucl.\ Phys.\  {\bf B831}, 133-161 (2010).
  [arXiv:0910.1100 [hep-ph]].

\bibitem{Griest:1990kh}
  K.~Griest and D.~Seckel,
  Phys.\ Rev.\  {\bf D43}, 3191-3203 (1991).
   
\bibitem{Karsch:2000kv}
  F.~Karsch, E.~Laermann, A.~Peikert,
  Nucl.\ Phys.\  {\bf B605}, 579-599 (2001) 
  [hep-lat/0012023].

\bibitem{Savage:2010tg}
  C.~Savage, G.~Gelmini and P.~Gondolo {\it et al.},
  [arXiv:1006.0972 [astro-ph.CO]].

\bibitem{Sorensen:2010hq}
  P.~Sorensen,
  JCAP {\bf 1009}, 033 (2010) 
  [arXiv:1007.3549 [astro-ph.IM]].

\end{thebibliography}
\end{document}